\documentstyle[12pt,graphicx]{article}
\begin{document}
\baselineskip 0.6cm

\def\simgt{\mathrel{\lower2.5pt\vbox{\lineskip=0pt\baselineskip=0pt
           \hbox{$>$}\hbox{$\sim$}}}}
\def\simlt{\mathrel{\lower2.5pt\vbox{\lineskip=0pt\baselineskip=0pt
           \hbox{$<$}\hbox{$\sim$}}}}

\begin{titlepage}

\begin{flushright}
UCB-PTH 04/16 \\
LBNL-55248 \\
\end{flushright}

\vskip 2.0cm

\begin{center}

{\Large \bf 
Acceleressence: Dark Energy from a Phase Transition \\ at the Seesaw Scale 
}

\vskip 1.0cm

{\large
Z. Chacko, Lawrence J. Hall, and Yasunori Nomura
}

\vskip 0.4cm

{\it Department of Physics, University of California,
                Berkeley, CA 94720} \\
{\it Theoretical Physics Group, Lawrence Berkeley National Laboratory,
                Berkeley, CA 94720}

\vskip 1.2cm

\abstract{
Simple models are constructed for ``acceleressence'' dark energy: the latent heat 
of a phase transition occurring in a hidden sector governed by the seesaw mass 
scale $v^2/M_{\rm Pl}$, where $v$ is the electroweak scale and $M_{\rm Pl}$ 
the gravitational mass scale.  In our models, the seesaw scale is stabilized 
by supersymmetry, implying that the LHC must discover superpartners with 
a spectrum that reflects a low scale of fundamental supersymmetry breaking. 
Newtonian gravity may be modified by effects arising from the exchange of fields 
in the acceleressence sector whose Compton wavelengths are typically of order 
the millimeter scale.  There are two classes of models.  In the first class the 
universe is presently in a metastable vacuum and will continue to inflate until 
tunneling processes eventually induce a first order transition.  In the simplest 
such model, the range of the new force is bounded to be larger than $25~{\rm \mu m}$ 
in the absence of fine-tuning of parameters, and for couplings of order unity it 
is expected to be $\approx 100~{\rm \mu m}$.  In the second class of models thermal 
effects maintain the present vacuum energy of the universe, but on further cooling, 
the universe will ``soon'' smoothly relax to a matter dominated era.  In this case, 
the range of the new force is also expected to be of order the millimeter scale 
or larger, although its strength is uncertain.  A firm prediction of this class of 
models is the existence of additional energy density in radiation at the eV era, 
which can potentially be probed in precision measurements of the cosmic microwave 
background.  An interesting possibility is that the transition towards a matter 
dominated era has occurred in the very recent past, with the consequence that 
the universe is currently decelerating.
}

\end{center}
\end{titlepage}

\section{Dark Energy from a Phase Transition}

Cosmological observations of Type~Ia supernovae, the cosmic microwave 
background radiation and large scale structure provide strong evidence that 
the universe is flat and composed of about 70\% dark energy and 30\% dark 
matter~\cite{Perlmutter:1998np,Spergel:2003cb,Tegmark:2003ud}.  The dark energy, 
which is driving a recent acceleration in the expansion in the universe, has 
negative pressure and cannot be interpreted as matter or radiation.  Rather, 
this unusual fluid may be some form of vacuum energy, with an energy density 
of order $(10^{-3}~{\rm eV})^4$.  A crucial question is whether this vacuum 
energy is time independent -- a ``hard'' cosmological constant, $\Lambda$ -- 
or evolves with time -- a ``soft'' vacuum energy.  An example of the latter is 
``quintessence'', a scalar field energy that evolves slowly over many decades 
of expansion of the universe~\cite{Peebles:1987ek}.  However, theories of 
quintessence involve an unnaturally small mass scale of order the Hubble 
parameter, $10^{-33}~{\rm eV}$, and do not explain why this field energy 
is just dominating the universe in the present epoch.  A hard cosmological 
constant also suffers from this ``Why now?'' problem; are we really witnessing 
the transition to an era of eternal inflation?

Our present understanding of the hot big bang is one of a succession of phase 
transitions interspersed with eras of smooth cooling.  The phase transitions 
are the cosmological manifestation of symmetry breaking, as the underlying 
vacuum shifts from one stable minimum to another.  Given the standard model of 
particle physics, it is extremely plausible that phase transitions occurred both 
as the temperature cooled through the electroweak scale, $v$, and through the 
scale of strong interactions, $\Lambda_{\rm QCD}$.  At higher temperatures there 
may well have been other phase transitions associated, for example, with the 
breaking of lepton number (to generate right-handed neutrino masses and for 
leptogenesis), the breaking of grand unified gauge symmetries, and the generation 
of an early era of cosmic inflation.  At each of these phase transitions it is 
likely that the universe was dominated for a period by the vacuum energy, or latent 
heat, of the associated change in vacuum state.  It therefore seems plausible to 
us that the present cosmic expansion is fueled by the soft vacuum energy of some 
phase transition associated with an energy scale of $10^{-3}~{\rm eV}$.  We label 
this phenomenon acceleressence. 

At first sight it does not seem reasonable that there could be a phase transition 
in the universe with a vacuum energy of order $(10^{-3}~{\rm eV})^4$, because we 
have not discovered any particle physics symmetry breaking at the $10^{-3}~{\rm eV}$ 
scale.  However, at low energies we know that interactions between particles can 
get very small, suppressed by inverse powers of a large mass scale, so that this 
new particle physics may be decoupled from us. For example, all interactions of 
the neutrino decouple at low energies, and its mass is often assumed to arise 
from inverse powers of the large right-handed neutrino mass $M_R$, $m_\nu \simeq 
v^2 / M_R$.  Suppose that acceleressence occurs in some hidden sector that 
interacts with the standard model only by gravity.  It could be that the mass 
scale of this sector is also generated by a seesaw mechanism, taking the value 
$v^2/M_{\rm Pl} \simeq 10^{-3}~{\rm eV}$, where the Planck mass, $M_{\rm Pl}$, 
is the scale of gravity.  It is intriguing that this ratio of known scales gives 
the observed energy scale of dark energy.  If such a sector underwent a phase 
transition it could cause cosmic acceleration, naturally explaining the ``Why now?'' 
problem~\cite{Arkani-Hamed:2000tc}.  In this paper we aim to construct the simplest 
models of such a hidden phase transition.

If there is to be a new scale of particle physics at $v^2/M_{\rm Pl} \simeq 
10^{-3}~{\rm eV}$, how can it be made stable to radiative corrections?  This 
appears even more daunting than the usual hierarchy problem of making $v$ stable 
to radiative corrections.  Remarkably supersymmetry can do both.  The usual 
hierarchy problem is solved by requiring that the scale of supersymmetry breaking 
in the standard model sector is of order the electroweak scale: $\tilde{m} \simeq v$. 
If this is the only breaking of supersymmetry in nature, and if the acceleressence 
sector only couples gravitationally to this supersymmetry breaking, then the scale 
of supersymmetry breaking in the acceleressence sector will naturally be 
\begin{equation}
  m_D \simeq \frac{\tilde{m}^2}{M_{\rm Pl}} 
    \simeq \frac{v^2}{M_{\rm Pl}} \simeq 10^{-3}~{\rm eV}, 
\label{eq:mD}
\end{equation}
and is stable to radiative corrections. 

Our task in this letter is to build the simplest models of acceleressence and
study their consequences. In general these theories possess a sector involving 
a scalar field $\phi$, the acceleressence field, such that all the supersymmetry 
breaking mass parameters in its potential are of order $m_D$.  We assume that 
the cosmological constant vanishes in the true minimum of the zero temperature 
potential. Before the phase transition associated with dark energy has occurred, 
$\langle \phi \rangle = 0$.  Clearly the crucial question is the behavior of the 
field near the origin.  In our first model, $\phi$ has a positive mass squared 
and we live in a false vacuum, with the phase transition induced by a trilinear 
$\phi^3$ interaction.  The tunneling rate to the true minimum may be very slow, 
so that the universe may enter a prolonged era of inflation before undergoing 
a first order phase transition.  In the second model, $\phi$ has a negative 
mass squared so that there is no local minimum of the potential at the origin. 
Nevertheless, thermal corrections to the effective potential are sufficient to 
maintain $\langle \phi \rangle = 0$ today.  This thermal barrier will rapidly 
disappear as the universe cools, and hence a phase transition towards a matter 
dominated era is imminent.  An interesting possibility is that this transition 
has already occurred, albeit in the very recent past.  In this case the universe 
is currently decelerating, leading to observable consequences for future precision 
measurements of the distance-redshift relation.

There are alternative ideas for understanding the size of the dark energy 
density.  It may be related to the neutrino energy density~\cite{Fardon:2003eh} 
in such a way that the dark energy resides in a slowly evolving scalar field. 
In axion models, the dark energy may be false vacuum energy associated with 
the size of explicit $U(1)_{\rm PQ}$ breaking coming from higher dimension 
operators~\cite{Barr:2001vh}.  Alternatively, in theories with extra spatial 
dimensions, the smallness of the dark energy density may follow from an 
exponential wavefunction profile in the bulk~\cite{Huber:2002np}.

\section{Models of Acceleressence}
\label{sec:models}

Our models have the following basic structure.  We have a sector that contains 
a scalar field $\phi$, the acceleressence field, responsible for the dark energy. 
This acceleressence sector interacts with the other sectors only through 
gravitationally suppressed interactions.  In particular, once supersymmetry is 
broken at the TeV scale in the standard model sector, its effects are transmitted 
to the acceleressence sector through the following operators:
\begin{equation}
  \int\! d^4 \theta\; \frac{X^{\dagger}X}{{M_{\rm Pl}}^2}
    {\Phi^{\dagger} \Phi},
\qquad
  \int\! d^4 \theta\; \frac{X+X^{\dagger}}{{M_{\rm Pl}}}{\Phi^{\dagger}\Phi},
\label{eq:op}
\end{equation}
where $\Phi$ is a chiral superfield containing $\phi$ as the lowest component, 
and $X$ is a superfield that breaks supersymmetry so that $F_X \simeq ({\rm TeV})^2$. 
This generates a soft supersymmetry-breaking mass and trilinear interaction 
for $\phi$ of order $F_X/M_{\rm Pl} \equiv m_D \simeq 10^{-3}~{\rm eV}$, which 
eventually produces dark energy of the observed size.  In theories where tree-level 
couplings between $\Phi$ and $X$ are absent, the operators in Eq.~(\ref{eq:op}) 
are generated effectively at loop level so that the scale of the fundamental 
supersymmetry breaking can be $F_X \simeq (10\!\sim\!100~{\rm TeV})^2$.

\subsection{Acceleressence from a false vacuum}
\label{subsec:model-1}

We now present the simplest model realizing our scenario. The acceleressence 
sector consists of a single chiral superfield $\Phi$ with the superpotential 
\begin{equation}
  W = \frac{\lambda}{3} \Phi^3.
\label{eq:model1-superpot}
\end{equation}
Taking supersymmetry breaking effects into account, the scalar potential 
is given by
\begin{equation}
  V = \lambda^2 |\phi|^4 - (A \phi^3 + {\rm h.c.}) + m^2 |\phi|^2 + V_0,
\label{eq:model1-scalarpot}
\end{equation}
where, without loss of generality, $\lambda$ and $A$ can be taken real and positive 
by rotating phases of fields.  Here, $m$ and $A$ are supersymmetry breaking 
parameters of order $m_D$, and $V_0$ is a constant determined by the condition 
that the cosmological constant is vanishing at the true minimum of the potential. 

We assume that $m^2$ is positive so that the model has a (local) minimum at the 
origin in field space.  If $A$ is sufficiently large, $9 A^2 > 8 \lambda^2 m^2$, 
the model has a second minimum at $\langle \phi \rangle \neq 0$.  This second 
minimum has a lower energy than the minimum at the origin if
\begin{equation}
  A > \lambda\, m.
\label{eq:model1-cond}
\end{equation}
We require this condition to be satisfied, so that the minimum at $\langle \phi 
\rangle = 0$ is only a local minimum.  Then, for $A$ sufficiently larger than 
$\lambda m$, we find 
\begin{equation}
  V_0 \simeq \frac{27A^4}{16\lambda^6} = O\Bigl(\frac{A^4}{\lambda^6}\Bigr).
\end{equation}
This is the vacuum energy density we observe today, if the $\phi$ field is trapped 
in the local minimum at the origin.  The trapping at the origin naturally occurs 
because it is likely that the universe starts at $\langle \phi \rangle = 0$ due to 
thermal effects or an induced $\phi$ mass term during inflation.  The lifetime for 
the decay of this metastable vacuum can easily be made longer than the age of the 
universe.  We then obtain the observed magnitude of the dark energy for natural 
values of parameters, $\lambda \simeq 1$ and $A \simeq 10^{-3}~{\rm eV}$.

\subsection{Acceleressence from a thermal vacuum}
\label{subsec:model-2}

In the model discussed above, the mass squared for the acceleressence field $\phi$ 
was assumed to be positive.  We can also consider a model in which the acceleressence 
field has a negative mass squared, but is trapped at the origin by thermal effects. 
Building such a model, however, is not entirely trivial due to the potential conflict 
between the observed size of the dark energy and the constraint from big-bang 
nucleosynthesis on the thermal energy density of the acceleressence sector.  
For example, we cannot simply take the model of section~\ref{subsec:model-1} 
and make $m^2$ negative, as the resulting model does not have a viable parameter 
region explaining the observed dark energy while satisfying all phenomenological 
constraints.  In this section we present a realistic model with the acceleressence 
field having a negative mass squared at the origin.

We take our acceleressence sector to be a supersymmetric $U(1)$ gauge theory with 
three chiral superfields $\Phi(+1)$, $\bar{\Phi}(-1)$ and $S(0)$, where the numbers 
in parentheses represent the $U(1)$ charges. The superpotential of the model is 
\begin{equation}
  W = \lambda S \Phi \bar{\Phi},
\label{eq:model2-superpot}
\end{equation}
where $\lambda$ is a coupling constant.  Now, suppose that $\Phi$ and $\bar{\Phi}$ 
obtain negative squared masses, $-m_{\phi}^2$ and $-m_{\bar{\phi}}^2$, and $S$ 
obtains a positive squared mass, $m_{s}^2$, from supersymmetry breaking (where 
$m_{\phi}^2, m_{\bar{\phi}}^2, m_{s}^2 > 0$ are of order $m_D^2$).  The scalar 
potential is given by 
\begin{eqnarray}
  V &=& |\lambda \phi \bar{\phi}|^2 
    + |\lambda s \phi|^2 + |\lambda s \bar{\phi}|^2
    + \frac{g^2}{2}(|\phi|^2-|\bar{\phi}|^2)^2 
\nonumber\\
  && + m_{s}^2 |s|^2 
    - m_{\phi}^2 |\phi|^2 - m_{\bar{\phi}}^2 |\bar{\phi}|^2 
    + V_0,
\label{eq:model2-scalarpot}
\end{eqnarray}
where $g$ is the $U(1)$ gauge coupling and $V_0$ is a constant to be chosen to 
make the cosmological constant vanishing at the true minimum of the potential. 
Note that in addition to the gauge symmetry, this theory possesses a global $U(1)$ 
symmetry under which $\phi$ and $\bar{\phi}$ have the same charge.  Here we have 
assumed the absence of scalar trilinear interactions for simplicity. 

For a somewhat suppressed superpotential coupling $\lambda^2 \ll g^2$, the minimum 
of the potential Eq.~(\ref{eq:model2-scalarpot}) lies at $\langle s \rangle = 0$ 
and $\langle \phi \rangle^2 \simeq \langle \bar{\phi} \rangle^2 \simeq (m_{\phi}^2 
+ m_{\bar{\phi}}^2)/2\lambda^2$.  At this point in field space both the gauge and 
the global symmetries are broken, so the spectrum contains a massless Goldstone 
boson.  Requiring $V=0$ at the minimum, we obtain
\begin{equation}
  V_0 \simeq \frac{(m_{\phi}^2 + m_{\bar{\phi}}^2)^2}{4\lambda^2}.
\label{eq:model2-V0}
\end{equation}
The potential Eq.~(\ref{eq:model2-scalarpot}) with Eq.~(\ref{eq:model2-V0}) does 
not support a constant vacuum energy, as there is no local minimum in the potential. 
The situation, however, can be different if this sector has a finite temperature 
$T \neq 0$.  In this case the effective potential receives an additional 
contribution, which is given at high temperature by
\begin{eqnarray}
  \delta V \simeq \frac{\lambda^2}{4} T^2 |s|^2 
    + \frac{g^2}{2} T^2 (|\phi|^2 + |\bar{\phi}|^2),
\label{eq:model2-thermalpot}
\end{eqnarray}
for $\lambda^2 \ll g^2$.  Therefore, as long as $g^2 T^2/2 \simgt m_{\phi}^2$ 
and $m_{\bar{\phi}}^2$, the fields are thermally trapped to $\langle s \rangle 
= \langle \phi \rangle = \langle \bar{\phi} \rangle = 0$ and the vacuum energy 
is given by $V_0$ in Eq.~(\ref{eq:model2-V0}).  This leads to the accelerated 
expansion of our universe as long as the thermal energy density is smaller than 
the vacuum energy density, which is actually the case as we will see below. 

In general, models of acceleressence using thermal effects are subject to 
severe phenomenological constraints.  The success of big-bang nucleosynthesis 
constrains the radiation energy density in the acceleressence sector, $\rho_\phi$, 
to be much smaller than that in photons, $\rho_\gamma$.  We here parameterize 
this constraint as $\rho_\phi \simlt \epsilon\, \rho_\gamma$, where $\epsilon 
\simeq 0.1$.  Since $\rho_\phi \simeq (\pi^2/30)g_\phi T^4$ and $\rho_\gamma 
\simeq (\pi^2/15) T_\gamma^4$, the constraint can be written as
\begin{equation}
  T^4 \simlt \frac{2\epsilon}{g_\phi} T_\gamma^4,
\label{model2-const1}
\end{equation}
where $g_\phi$ is the number of effective degrees of freedom in the acceleressence 
sector and $T_\gamma$ is the photon temperature.  In the present model, $g_\phi 
= 15$.  The temperature of the acceleressence sector must also satisfy the condition 
such that it traps the fields in a local minimum.  In the present model, this 
condition is given by
\begin{equation}
  \frac{g^2}{2} T^2 \simgt m_{\phi}^2,\, m_{\bar{\phi}}^2.
\label{model2-const2}
\end{equation}
Using Eqs.~(\ref{model2-const1},~\ref{model2-const2}) in Eq.~(\ref{eq:model2-V0}) 
we then find that the parameters of the model must satisfy
\begin{equation}
  \frac{\lambda}{g^2} \simlt \left( \frac{\epsilon}{30} 
    \frac{T_\gamma^4}{V_0} \right)^{\frac{1}{2}} \simeq 10^{-3},
\label{model2-modelpara}
\end{equation}
where we have used $V_0 \simeq 3 \times 10^{-11}~{\rm eV}^4$ and $T_\gamma \simeq 
2.4 \times 10^{-4}~{\rm eV}$.  Assuming $g = O(1)$, this requires a somewhat small 
coupling of $\lambda \simlt 10^{-3}$.  The small value of the quartic coupling for 
the acceleressence field is protected against radiative corrections by imposing an 
approximate $U(1)_R$ symmetry on the acceleressence sector that forbids a $U(1)$ 
gaugino mass.  Note that with the condition Eq.~(\ref{model2-const1}) satisfied, 
the radiation energy density in the acceleressence sector, $\rho_\phi$, is much 
smaller than the vacuum energy density, $V_0$, so that the accelerated expansion 
of the universe necessarily follows. 

What is the future of the universe in such a scenario?  At some point the
temperature in the acceleressence sector falls to a point where the inequality 
Eq.~(\ref{model2-const2}) is no longer satisfied.  The field of higher $m^2$ 
then start evolving at the Hubble rate, tracking the minimum of its thermal 
effective potential.  However, during this era the vacuum energy only changes 
by a negligible amount so that $w$ is still essentially $-1$.  This era ends 
when the field reaches some critical value.  At this point thermal effects 
in the potential disappear, and both $\phi$ and $\bar{\phi}$ fields rapidly 
acquire large values and oscillate about the minimum of the potential.  The 
vacuum energy is then essentially instantly converted first into matter and 
then into the radiation energy of Goldstone bosons in the acceleressence sector. 
This makes the ultimate future of the universe to be dominated by cold dark 
matter.  An intriguing possibility is that the conversion of dark energy to 
radiation energy described above has already occurred, albeit in the very 
recent past, with the deceleration parameter jumping from $\approx -0.5$ to 
$\approx +0.8$, so that the universe is currently decelerating.  The present 
value of the deceleration parameter may vary as the model is changed and typically 
lies between $\approx +0.5$ and $\approx +0.8$ in simple models.  This leads 
to observable consequences for the very recent expansion of the universe, which 
may be probed by future observations of the distance-redshift relation.

\subsection{Origin of supersymmetry breaking}
\label{subsec:origin}

In the models discussed above, the scale of the fundamental supersymmetry 
breaking should be low, $F_X \simeq (1\!\sim\!100~{\rm TeV})^2$, to have 
$m_D \simeq 10^{-3}~{\rm eV}$.  Here we discuss some explicit examples for such 
theories.  The first example we discuss is a class of theories where supersymmetry 
is dynamically broken at around a TeV by nearly conformal gauge interactions. 
These theories have a dual 5D description in which supersymmetry is broken on 
the infrared (TeV) brane in a warped extra dimension~\cite{Gherghetta:2000qt,%
Nomura:2003qb}.  The standard-model gauge fields propagate in the bulk and 
matter fields are located on the ultraviolet (Planck) brane.  Now, suppose 
we introduce a $\Phi$ superfield on the Planck brane with the superpotential 
of Eq.~(\ref{eq:model1-superpot}).  Supersymmetry breaking on the TeV brane then 
produces the soft supersymmetry breaking parameters for $\phi$ through anomaly 
mediation~\cite{Randall:1998uk}: $m^2 = 4\lambda^4 M^2$ and $A = 2\lambda^3 M$, 
where $M = m_{3/2}/16\pi^2$ with $m_{3/2}$ the gravitino mass.  The soft mass 
squared $m^2$ also receives contribution from the possible terms on the Planck 
brane: $\int\! d^4\theta \, H^\dagger H \Phi^\dagger \Phi/M_{\rm Pl}^2$ if the 
Higgs fields are localized towards the Planck brane ($c_H > 1/2$), or $\int\! 
d^2\theta \, (\Phi/M_{\rm Pl}) W^\alpha W_\alpha$ via a finite sum of one-loop 
diagrams involving the standard-model gauge bosons and their superpartners. 
In this class of models $m_{3/2}$ can naturally be in the $10~{\rm TeV}$ 
region~\cite{Nomura:2003qb}, so that the desired values for $m^2$ and $A$ 
(satisfying Eq.~(\ref{eq:model1-cond})) can be generated in certain parameter 
regions.  This then gives the correct amount of dark energy with $\lambda 
= O(0.1\!\sim\!1)$.

Another example is based on the gauge mediation scenario~\cite{Dine:1981gu}, but 
with the messenger and standard-model matter fields geometrically separated by 
extra spatial dimensions, for example by a flat extra dimension having the size 
around the grand unification scale.  The standard-model gauge fields located 
in the bulk then transmit supersymmetry breaking from the messenger sector to 
the standard model sector.  The size of supersymmetry breaking in the messenger 
sector can be of order $10\!\sim\!100~{\rm TeV}$.  If this is the largest 
supersymmetry breaking in the model, the $\Phi$ field located on the standard-model 
brane receives soft supersymmetry-breaking parameters $m^2$ and $A$ with the 
appropriate size through anomaly mediation and operators located on that brane, 
as in the models in the previous paragraph.  This type of model also provides 
an explicit example of our scenario.

In both examples of supersymmetry breaking given above, the spatial separation 
of $\Phi$ from the supersymmetry breaking field $F_X$ implies that $\phi$ does 
not feel supersymmetry breaking from tree-level supergravity mediation; rather 
the dominant contribution arises at one loop from anomaly mediation.  This leads 
to $m \simeq 10^{-4} \lambda^2 (\sqrt{F_X}/10~{\rm TeV})~{\rm eV}$, easily allowing 
$F_X$ to be in the range of $10\!\sim\!100~{\rm TeV}$.  These low values of $F_X$ 
imply a rather light gravitino mass of $0.01\!\sim1~{\rm eV}$, so that the lightest 
supersymmetric particle cannot be weakly interacting cold dark matter.  The dark 
matter in our theories should be provided by a generic particle with TeV-sized 
mass and cross section.  Such a particle may arise from fields localized on the 
infrared brane in warped models.

\section{Potential Signatures}
\label{sec:signals}

Here we discuss some potential signatures of our models.  In our scenario, fields 
in the acceleressence sector may interact with the standard model fields through 
Planck-suppressed operators.  Suppose that the $\Phi$ field in the model of 
section~\ref{subsec:model-1} interacts with the standard-model gauge fields 
through the following operator:
\begin{equation}
  \int\! d^2\theta \; \frac{\Phi}{M_{\rm Pl}}
    {\rm Tr}\left[{W^{\alpha} W_{\alpha}}\right]
  \rightarrow \frac{\phi}{M_{\rm Pl}}
    {\rm Tr} \left[F^{\mu \nu} F_{\mu \nu}\right],
\label{eq:long-range}
\end{equation}
where $W_\alpha$ represents the field-strength superfields for the standard-model 
gauge fields.  This induces a modification of the gravitational potential between 
two bodies of masses $m_1$ and $m_2$ separated at a distance $r$ through the $\phi$ 
exchange: $V_{\rm grav} = -(1 + \alpha e^{-r/l}) G_N m_1 m_2/r$.  The size of the 
modification, $\alpha$, depends on an unknown coefficient of the operator in 
Eq.~(\ref{eq:long-range}); we naturally expect that it is of order unity, but 
it could be small.  The range of modification, $l$, is determined by the mass 
of the $\phi$ scalar: $l \simeq m^{-1}$.  An interesting aspect of the model 
is that the size of the dark energy has an implication for the range of the 
modification $l$.  To see this we can explicitly minimize the potential of 
Eq.~(\ref{eq:model1-scalarpot}) and write $V_0$ as
\begin{equation}
  V_0 = \frac{m^4}{\lambda^2} f \left( \frac{\lambda m}{A} \right),
\end{equation}
where $f$ is a function defined by $f(x)=(27-36x^2+8x^4+(9-8x^2)^{3/2})/32x^4$. 
The function $f(x)$ has the property that for $x < 0.8$, $f(x) > 1$.  Therefore, 
for a parameter region $\lambda m/A \simlt 0.8$ we obtain a lower bound on 
$l \simeq m^{-1}$:
\begin{equation}
  l \simgt \frac{1}{\sqrt{\lambda}} {V_0}^{-\frac{1}{4}}.
\label{eq:bound}
\end{equation}
For the vacuum at $\phi = 0$ to be metastable, $\lambda m/A < 1$ (see 
Eq.~(\ref{eq:model1-cond})), so that this bound on $l$ applies to most of 
the parameter space of the model. (The exception is when $A$ is very close 
to $\lambda m$.)

A numerical bound on the range $l$ is obtained from the upper bound on the 
coupling $\lambda$: $\lambda \simlt 4\pi$.  Using $V_0 \simeq 3 \times 
10^{-11}~{\rm eV}^4$, we obtain $l \simgt 24~{\rm \mu m}$.  An even stronger 
bound arises if we require that the coupling $\lambda$ is perturbative up to the 
Planck scale.  In this case the renormalization group analysis gives that $\lambda 
\simlt 0.76$ at the scale $m_D$, so that we obtain $l \simgt 110~{\rm \mu m}$. 
These distance scales are within striking range of experiments searching for 
deviations from Newtonian gravity at sub-millimeter distances~\cite{Hoyle:2000cv}. 
For a review of the current and future experimental status, and for alternative 
theories which also predict deviations from Newtonian gravity at sub-millimeter 
distances, see~\cite{Long:2003ta}.

A similar signature can be obtained in the model of section~\ref{subsec:model-2} 
through the coupling of the $S$ field to the standard-model gauge fields 
(Eq.~(\ref{eq:long-range}) with $\Phi$ replaced by $S$ -- operators linear in 
$\Phi$ or $\bar{\Phi}$ are forbidden by the gauge symmetry).  In this case $l$ 
is determined by the mass of the scalar $s$: $l \simeq m_s^{-1}$, which does 
not have a solid bound as in the previous case.  However, the naturalness of the 
model implies that radiative corrections to $m_{\phi}$ and $m_{\bar{\phi}}$ from 
$m_s$ cannot be much larger than the values of $m_{\phi}$ and $m_{\bar{\phi}}$ 
themselves, which gives the bound $l \simgt (\lambda/4\pi) \sqrt{\ln(M_*/\Lambda)}\, 
m_{\phi}^{-1}$.  For $\lambda \simeq 10^{-3}$, this gives $l \simgt 2~{\rm \mu m}$. 
Moreover, in the case that all the supersymmetry-breaking masses are the same 
order, i.e. $m_{s} \sim m_{\phi}, m_{\bar{\phi}}$, we obtain much tighter bound 
$l \simgt 3~{\rm mm}$.  The strength of the modification, $\alpha$, can be of 
order $1$ but, if the $S$ field is responsible for the suppression of $\lambda$ 
in Eq.~(\ref{eq:model2-superpot}), it could be of order $\lambda^2 \simeq 10^{-6}$.

Another possible signature of our models arises from the radiation energy density 
in the acceleressence sector, which we call dark radiation.  This is especially 
interesting in the model of section~\ref{subsec:model-2} because the acceleressence 
sector necessarily has a finite temperature. In particular, if the bound on 
Eq.~(\ref{model2-const1}) is nearly saturated, which is the case if $\lambda$ 
is not much smaller than $10^{-3}g^2$, the radiation energy density $\rho_\phi$ 
is close to the upper bound allowed by nucleosynthesis, implying that this dark 
radiation will be seen in future cosmic microwave background experiments such as 
PLANCK or CMB-Pol.  The signature from the dark radiation could also arise in the 
model of section~\ref{subsec:model-1} if the temperature of the acceleressence 
sector is not much lower than the value allowed by the big-bang nucleosynthesis 
constraint.

Finally, we note that the equation of state for acceleressence is essentially 
$w = -1$, except perhaps in the very recent past.  Furthermore, superpartners 
must be discovered at the LHC with a spectrum that reflects a low mediation 
scale for supersymmetry breaking.  If either of these proves to be incorrect, 
acceleressence dark energy will be excluded, at least in its minimal form as 
described in this paper.

\section*{Acknowledgments}

This work was supported in part by the Director, Office of Science, Office 
of High Energy and Nuclear Physics, of the US Department of Energy under 
Contract DE-AC03-76SF00098 and DE-FG03-91ER-40676, and in part by the 
National Science Foundation under grant PHY-00-98840.

\end{document}